\documentclass[12pt]{article}
\newcommand{\be}{\begin{equation}}
\newcommand{\ee}{\end{equation}}
\newcommand{\bc}{\begin{center}}
\newcommand{\ec}{\end{center}}
\newcommand{\ber}{\begin{eqnarray}}
\newcommand{\ear}{\end{eqnarray}}
\newcommand{\ba}{\begin{array}}
\newcommand{\ea}{\end{array}}

\newcommand{\al}{\alpha}
\newcommand{\ar}{{\cal A}}
\newcommand{\bt}{\beta}

\newcommand{\de}{\delta}

\newcommand{\ds}{\rm{d}\sigma}
\newcommand{\Ds}{\frac{\rm{D}}{\rm{d}\sigma}}
\newcommand{\dt}{\rm{d}\tau}
\newcommand{\Dt}{\frac{\rm{D}}{\rm{d}\tau}}
\newcommand{\el}{\ell}

\newcommand{\et}{\eta}
\newcommand{\ff}{\aleph}
\newcommand{\fr}{\frac}
\newcommand{\Ga}{\Gamma}
\newcommand{\ga}{\gamma}

\newcommand{\la}{\lambda}
\newcommand{\lb}{\label}

\newcommand{\Lf}{{\cal L}_\aleph}
\newcommand{\Lg}{{\cal L}}
\newcommand{\Ls}{\check{\cal L}}
\newcommand{\n}{\nonumber\\}

\newcommand{\na}{\nabla}

\newcommand{\om}{\omega}
\newcommand{\p}{\partial}

\newcommand{\pr}{{\cal P}}
\newcommand{\rh}{\rho}

\newcommand{\si}{\sigma}
\newcommand{\sq}{\sqrt}
\newcommand{\st}{\stackrel}
\newcommand{\ta}{\tau}

\newcommand{\Te}{\Theta}

\begin{document}
\title{The Rotation and Shear of a String.}
\author{Mark D. Roberts,
54  Grantley Avenue,  Wonersh Park GU5 OQN,\\
Email: mdrobertsza@yahoo.co.uk,  
http://cosmology.mth.uct.ac.za/$\sim$~roberts}
\maketitle
\bc Eprint:  hep-th/0204236  \ec
\bc Comments:  17 pages,  
Version 2: result generalized and contact established with the approach of others.
Version 3: some sub/superscript corrections on the internal labels of the Nambu-Goto string.\ec
\begin{abstract}
Whether a string has rotation and shear can be investigated 
by an anology with a congruence of point particles.
Rotation and shear involve first covariant spacetime derivatives of a vector field and,  
because the metric stress tensor for both the point particle 
and the string have no such derivatives,  
the best vector fields can be identified by requiring the conservation
of metric stress.
It is found that the best vector field is a non-unit accelerating field
in $x$,  rather than a unit non-accelerating vector 
involving the momenta;
it is also found that there is an equation obeyed by the spacetime derivative 
of the Lagrangian
$$\Lg^\mu=P^\mu_a(F\st{a}{\Te}+F_a)$$
using notation which will be defined in the paper.
The relationship betwen membranes and fluids is looked at,
and it is shown how to produce a membrane with arbitrary $\Ga$
for the $\Ga$-equation of state.
\end{abstract}
{\small\tableofcontents}
\section{Introduction}
\lb{intro}
In general relativity the covariant derivative of a vector field can be decomposed into parts
called the rotation,  shear and so forth,  
see for example \cite{HE} and the appendix \S\ref{app}:  
these reduce to the same objects in newtonian fluid dynamics.
This vector field can be thought of as being tangent to the world-line of a particle 
and so it is of particular interest what these objects are for a congruence of 
point particles whose motion has a lagrangian description.
From a point particles lagrangian one can produce a spacetime metric stress 
which is proportional to the projection tensor and fills the whole spacetime.
Clearly this stress cannot be of a single point particle,
which must have a $\de$-function stress to be point-like;
how such stresses can be produced is addressed in section \S6.
For the simplest point particle actions,  the spacetime metric stress cannot
represent interacting or self-interacting particles.
Particles and strings will self-interact via deviation equations,
see roberts \cite{mdr}.
As the spacetime metric stress fills a portion of the spacetime it is said to 
represent a congruence of particles,
by anology with a "congruence of curves" \cite{HE}p.69.
It is assumed that the best vector field describing a congruence of particles 
is also the best vector for an individual member of the congruence.
It is not immediate what the best vector field to choose to constuct rotation and shear is:
one could choose the unit vector field $P^\mu/m$,
which would considerably simplify subsequent calculations;
or one could choose the non-unit vector field $\dot{x}^\mu$,
which has acceleration.
Here it is suggested that the best choice is $\dot{x}^\mu$,
as the conservation of the spacetime metric stress tensor then takes the simplest form.
The analysis is then extended from the particles to strings and membranes.
Two topics not looked at are:
firstly {\it boundary conditions},
in fact only the rotation and shear of a part of the string are looked at,
whether the whole object has these properties is left open;
secondly its {\it classification},
the existence,  or otherwise,  of shear and rotation,
can signify the classification,  for example Petrov type,
of the spacetime under consideration;
this can be thought of as a classification of the covariant derivatives of the vector field:
thus in the present case there could be a 
classification of the spacetime covariant derivative
and the internal covariant derivative,
and perhaps a mixture of the two.
The treatment of the spacetime metric stress of a congruence of particles is
generalized to the Nambu-Goto action in section \S\ref{nga},
and for several membrane actions in section \S\ref{mem},
all the time the notation used is succesively generalized.
The spacetime metric stress of congruences of strings and membranes contains terms in both
the spacetime metric and projection tensor,  as does a perfect fluid.
The relationship between the two is examined,
the main difference being that membranes have much more complex projection tensors.
The notation used is that of \cite{HE}, except that
$p$ is the dimensions of the brane, $P$ denotes the momentum, and ${\pr}$ is the pressure, 
$\ga$ is the internal metric, $\Ga$ is used for the $\ga$-equation of state.
\section{A Congruence of Particles.}\lb{pp}
\subsection{The Point Particle's Action.}\lb{ppa}
The coordinate space action of a point particle is
\be
S=\int^{\ta_2}_{\ta_1}{\rm d}\ta\Lg,
\lb{2.1.1}
\ee
for the simplest case of a non-interacting particle,
the specific lagrangian is
\be
\Lg=-m\el,~~~
\el=\sq{-\dot{x}^2},~~~
\dot{x}^\al=\Dt x^\al.
\lb{2}
\ee
This is sometimes also called the non-linear form,  
and sometimes the determinant form.
Varying the velocity $\dot{x}^\al$ 
one can interchange the variation and the covariant derivative
\be
\de\dot{x}^\al
=\de\Dt x^\al
=\rm{D}\fr{\de x^\al}{\rm{d}\ta},
\lb{2.1.2}
\ee
So that the variation of the action is
\be
\de S=
\fr{\p\Lg}{\p \dot{x}^\mu}\de x^\mu |^{\ta_2}_{\ta_1}
-\int^{\ta_2}_{\ta_1}\rm{d}\ta \de x^\mu
\Dt
\fr{\p\Lg}{\p \dot{x}^\mu}.         
\lb{2.1.3}
\ee
The first term of the varied action \ref{2.1.3} 
can be made to vanish by using suitable boundary conditions,
a simple choice is
\be
\de x^\mu\fr{\p\Lg}{\p \dot{x}^\mu}|_{\ta_1}  
=\de x^\mu\fr{\p\Lg}{\p \dot{x}^\mu}|_{\ta_2}
=0,
\lb{2.1.4}
\ee
which are obeyed in the particular case
\be
\de x^\mu|_{\ta_1}=\de x^\mu|_{\ta_2}=0.
\lb{2.1.5}
\ee
The second term of the varied action \ref{2.1.3} 
vanishes when the equations of motion
\be
\Dt\fr{\p\Lg}{\p\dot{x}^\mu}=0,
\lb {2.1.6}
\ee
are obeyed.  Substituting $\p\Lg/\p\dot{x}^\mu$ 
for the particular Lagrangian defined in \ref{2} gives
\be
m\Dt\fr{\dot{x}^\mu}{\el}\st{\ref{2}}{=}0
\lb{2.1.7}
\ee
where in this and subsequent cases $\st{\ref{2}}{=}$
signifies that the equality holds only for 
a specific Lagrangian in this case \ref{2}.
The covariant nature of $\rm{D}/\rm{d}\ta$ 
ensures that \ref{2.1.7} is the geodesic equation.
\subsection{Introducing Momenta and the Hessian.}\lb{imh}
The momentum and its absolute derivative are defined by
\be
P_\mu
\equiv\fr{\de S}{\de \dot{x}^\mu}
=\fr{\p\Lg}{\p \dot{x}^\mu}
\st{\ref{2}}{=}m\fr{\dot{x}^\mu}{\el}.~~~~~~~~~~~
\dot{P}_\mu                        
=\Dt P_\mu,
\lb{2.1.8}
\ee
The boundary condition \ref{2.1.4} can be expressed in terms of the momenta
\be
\de x^\mu P_\mu|_{\ta_1}
=\de x^\mu P_\mu|_{\ta_2}
=0,
\lb{2.1.9}
\ee
as can the equation of motion
\be
\dot{P}^\mu=0.
\lb{2.1.10}
\ee
The Hessian is defined as
\be
W_{\mu\nu}
\equiv\fr{\de^2S}{\de\dot{x}^\mu\de\dot{x}^\nu}   
=\fr{\p^2\Lg}{\p\dot{x}^\mu\p\dot{x}^\nu}
\st{\ref{2}}{=}\fr{m}{\el}h_{\mu\nu}
\lb{2.1.10b}
\ee
$h_{\mu\nu}$ is the projection tensor defined below \ref{2.1.12}.
This equality suggests that the Hessian
can be thought of as a generalized projection.
For the particular Lagrangian \ref{2} 
the geodesic equation \ref{2.1.7} or \ref{2.1.10}
can be expressed in terms of the specific projection tensor
\be
\dot{P}^\mu
=W^{\mu\nu}\ddot{x}_\nu
\st{\ref{2}}{=}\fr{m}{\el}h^{\mu\nu}\ddot{x}_\nu=0,
\lb{2.1.11}
\ee
where the specific projection tensor familiar from general relativity is
\ber
h^{\mu\nu}
&\equiv& g^{\mu\nu}-\fr{\dot{x}^\mu\dot{x}^\nu}{\dot{x}^2}
=g^{\mu\nu}+\fr{P^\mu\dot{x}^\nu}{m\el}
=g^{\mu\nu}+\fr{P^\mu P^\nu}{m^2},\nonumber\\
h^{\mu\al}h_{\nu\al}&=&h^\nu_{.\nu}=g^{\mu\al}h_{\nu\al},\nonumber\\
h^{\mu\nu}\dot{x}_\nu&=&0,~~~
h=d-1
\lb{2.1.12}
\ear
\subsection{Some Standard Equations.}\lb{sse}
The spacetime metrical stress is
\be
T^{\mu\nu}=-\fr{2}{\sq{-g}}\fr{\de S}{\de g_{\mu\nu}}
          =-2\fr{\p \Lg{x}}{\p g_{\mu\nu}}+g^{\mu\nu}(x)\Lg(x),
\lb{2.1.22}
\ee
and it is often taken to be the energy-momentum tensor corresponding to $\Lg(x)$
It is assumed that this tensor is conserved
\be
T^{\mu\nu}_{.~.~;\nu}=0,
\lb{2.1.23}
\ee
For the point particle \ref{2} the energy-momentum tensor is
\be
T_{\mu\nu}
=-\el^2W_{\mu\nu}
\st{\ref{2}}{=}-m\el h_{\mu\nu}
\st{\ref{2}}{=}-m\el g_{\mu\nu}-P_\mu\dot{x}_\nu
\lb{2.1.27}
\ee
The perfect fluid energy-momentum tensor Eq.3.8\cite{HE} is
\be
T_{\mu\nu}=(\mu+\pr)V_\mu V_\nu+\pr~g_{\mu\nu},~~~
          =(\mu+\pr)h^{\mu\nu}-\mu g_{\mu\nu},~~~
V_\mu V^\mu=-1, 
\lb{2.1.27b}
\ee
in this case $h^{\mu\nu}$ is given by the first equality of \ref{2.1.12}
with $V^\mu=\dot{x}^\mu/\el$.
\ref{2.1.27b} can encompass the energy-momentum of many systems,
such as scalar fields,  essentially this is
because of the freedom to choose an equation of state.
When $\mu=0$ the perfect fluid \ref{2.1.27b} has energy-momentum tensor
\be
T_{\mu\nu}=\pr h_{\mu\nu},
\lb{2.1.27c}
\ee
comparing with the second equality of \ref{2.1.27} the energy-momentum of a point particle
can be thought of as a perfewct fluid with density $\mu=0$ and pressure
\be
\pr\st{\ref{2}}{=}-m\el.
\lb{1.1.27d}
\ee
From the properties of the projection tensor 
\ref{2.1.12}, \ref{2.1.27} gives
\be
\dot{x}^\nu T_{\mu\nu}\st{\ref{2}}{=}0,~~~
h^{\mu\al}T_{\al\nu}
\st{\ref{2}}{=}-m\el~h^\mu_{.~\nu}
\lb{2.1.28}
\ee
The first of these equations has as a consequence that one cannot
usefully take a definition such as
\be
P^\mu=V_\nu T^{\mu\nu},
\lb{2.1.29}
\ee
as taking $V^\nu$ proportional to $\dot{x}^\nu$ gives $P^\nu=0$.
\subsection{Derivatives of the Stress}\lb{dos}
Stress conservation \ref{2.1.23} is explicitly
\be
T^{\mu\nu}_{.~.~;\nu}
=-\fr{m}{\el}\left(\dot{x}^\al(\dot{x}^\mu_{.~\al}-\dot{x}^{~~\mu}_{\al.})
+\dot{x}^\mu\dot{x}^\al_{.~\al}
-\fr{\dot{x}^\al\dot{x}^\bt}{\dot{x}^2}\dot{x}_{\al\bt}\dot{x}^\mu\right),
\lb{2.1.30}
\ee
and is satisfied under the stringent condition
\be
\dot{x}^\al_{.~;\bt}=0.
\lb{2.1.31}
\ee
The equation \ref{2.1.31} can be thought of as
\be
\dot{x}^\al_{.~;\bt}
=\na_\bt\Dt x^\al               
=\Dt\na_\bt x^\al
=\Dt\left(\de^\al_\bt+\Ga^\al_{\bt\ga}x^\ga\right)
=\Dt\Ga^\al_{\bt\ga}x^\ga
\lb{2.1.32}
\ee
implying that either the metric is flat or the coordinates are special.
At first sight the first term in \ref{2.1.30} looks as if it is related 
to the vorticity and the second term to the expansion of $\dot{x}^\al$;
however $\dot{x}^\al$
is not of unit size {\it i.e.} $\dot{x}_\al\dot{x}^\al\ne1$,  
or even constant vector, so that the identification is not straightforward,  
and the expression appears not to simplify.  
Simpification is more easily achieved
with the mixed $x,P$ form of the stress,  
see the last equality of \ref{2.1.27} to get 
the conservation equation
\be
T^{\mu\nu}_{.~.~;\nu}
\st{\ref{2}}{=}-m\el^\mu-P^\mu_{.~;\nu}\dot{x}^\nu+P^\nu_.\dot{x}^\nu_{.~;\nu}
\lb{2.1.33}
\ee
Taking the absolute derivative to be given by
\be
\Dt=\dot{x}^\nu\na_\nu,
\lb{2.1.34}
\ee
although it could be a function times this,
the second term in \ref{2.1.33} is the $\ta$ covariant derivative of $P$,  
{\it i.e.}
\be
P^\mu_{.~\nu}\dot{x}^\nu=\dot{P}^\mu,
\lb{2.1.35}
\ee
and this vanishes by the equations of motion \ref{2.1.10}.
This term would still vanish if there was an additional function on the 
left of the right hand side of \ref{2.1.34}.
The last term in \ref{2.1.33} can be taken to be the expansion of $x$
\be
\dot{x}^\nu_{.~;\nu}=\st{x}{\Te}.      
\lb{2.1.36}                
\ee
Substituting \ref{2.1.10}, \ref{2.1.23}, \ref{2.1.35} \& \ref{2.1.36} 
into \ref{2.1.33} gives
\be
m\el^\mu=\st{x}{\Te}P^\mu,          
\lb{2.1.37}
\ee
which equates the expansion of $x$ to the change in the lenght $\el$.
Now 
\be
m\el^\mu=-\Lg^\mu,
\lb{2.1.37b}
\ee
so that
\be
-\Lg^\mu=\st{x}{\Te}P^\mu.
\lb{2.1.37c}
\ee
Expanding \ref{2.1.37} for $x$ gives
\be
\dot{x}^\al_{.\al}\dot{x}^\mu=-\dot{x}_\al\dot{x}^\al_{.\mu}
=-\ddot{x}^\mu+2\dot{x}^\al\dot{x}_{[\mu\al]},
\lb{2.1.37d}
\ee
which shows that the result \ref{2.1.27} 
could not be directly obtained by differentiating the 
Lagrangian,  neither does it follow immediateky from \ref{2.1.30}.
Collecting together \ref{2.1.37}, \ref{2.1.37c} and \ref{2.1.37d} gives
\be
\Lg^{;\mu}
=-m\el^{;\mu}
=\fr{m}{\el}\dot{x}^\al\dot{x}^\al_{.\mu}
=\fr{m}{\el}(\ddot{x}-2\dot{x}^\al\dot{x}_{[\mu\al]})
=-\fr{m}{\el}\dot{x}^\al_{.\al}\dot{x}^\mu
=-\st{x}{\Te}P^\mu.
\lb{2.1.37e}
\ee

The preceeding way of requiring stress conservation occures in two
parts seperately is not the same as the standard approach when a 
vector field is present,  for example for a perfect fluid.
There one requires $V_\al T^{\al\bt}_{.~.~;\bt}$
and $h^\al_{.~\bt}T^{\bt\ga}_{.~.~;\ga}$
to vanish seperately.  To see what happens if this is tried in the 
present case notice the form of stress conservation
\be
T^{\mu\nu}_{.~.~;\nu}
\st{\ref{2}}{=}-m\el_\nu h^{\mu\nu}_{.~.}-m\el h^{\mu\nu}_{.~.~;\nu}
\lb{2.1.38}
\ee
Transvecting with $V^\mu$ gives the second term equal to zero,
substituting back into \ref{2.1.38} 
gives the first term equal to zero,
so that $\el$ is a constant,
which implies a very restricted system.
The middle term of \ref{2.1.33} is going to vanish by the equations
of motion and \ref{2.1.37} is just what remains,
so that this method imposes no restriction on the system.
\subsection{The Phase Space Action.}\lb{psa}
The phase space action is
\be
S=\int^{\ta_2}_{\ta_1}\dt\Lg_{ph},~~~
\Lg_{ph}\st{\ref{2}}{=}\dot{x}\cdot P+\la_1\si^1.
\lb{2.1.44}
\ee
this is sometimes called the first order form,  
and sometimes the Hamiltonian form.
Varying with respect to $\la$, $P^\al$ and $x^\al$ gives
\ber
\fr{\de S}{\de \la_1}&=&P^2+m^2,~~~
\fr{\de S}{\de P_\al}=\dot{x}^\al+2\la_1 P^\al,\\
\int^{\ta_2}_{\ta_1}\dt P_\al\de\dot{x}^\al
&=&\int^{\ta_2}_{\ta_1}\dt\left[\Dt(P_\al\de x^\al)-\de x^\al\Dt P_\al\right]
=P_\al\de x^\al|^{\ta_2}_{\ta_1}
-\int^{\ta_2}_{\ta_1}\dt\de x^\al\dot{P}_\al,\nonumber
\lb{2.1.45}
\ear 
respectively.
The first equality of the $\de x^\al$ variation follows from \ref{2.1.2}.
The first term of the $\de x^\al$ variation is just the boundary condition \ref{2.1.3}
and so can be satisfied by \ref{2.1.9}.
The second term of the $\de x^\al$ variation gives the equation of motion \ref{2.1.10}.

At first sight the treatment of the energy-momentum is different from the
coordinate space approach.
The absense of the square root changes the first term by a factor of $2$
and the constraint gives an extra term so that by substituting
\be
T^{\mu\nu}=-2u^{[\mu}P^{\nu]}-2\la_1 P^\mu P^\nu+g^{\mu\nu}\Lg,
\lb{2.1.46}
\ee
and this is the same as before \ref{2.1.27}.
Note that the constraint contributes to the value of the stress \ref{2.1.46},
but not the Lagrangian \ref{2.1.44} 
\subsection{The Second Order Action}
\lb{soa}
Removing $P^\al$ from the Lagaingian \ref{2.1.44}
using $P^\mu=-x^\mu/2\la_1$ from \ref{2.1.45} the Lagrangian reduces to its second order form
\be
\Lg_2=\fr{1}{2}(\et^{-1}\dot{x}^2-\et m^2),~~~ 
2\la_1=-\et.
\lb{2.1.49}
\ee
Requiring that the variation of \ref{2.1.49} with respect $\et$ to vanish gives
\be
\et=\pm\fr{\sq{-\dot{x}^2}}{m},
\lb{n37}
\ee
requiring the variation of \ref{2.1.49} with repest to $\dot{x}^\mu$ to vanish gives
\be
\Dt\left(\fr{\dot{x}^\al}{\et}\right)=0,
\lb{n38}
\ee
substituting the value of $\eta$ \ref{n37} gives the equation of motion \ref{2.1.7}.
The energy-momentum tensor \ref{2.1.22} is
\be
T^{\mu\nu}=\fr{1}{\et}\left(
-\dot{x}^\mu\dot{x}^\nu+\fr{1}{2}g^{\mu\nu}(\dot{x}^2-m^2\et^2)\right),
\lb{n39}
\ee
substituting the value of $\et$ \ref{n37} back into \ref{n39} gives the stress \ref{2.1.27}.

\subsection{The Rotation and Shear of a Point Particle.}\lb{rspp}
The identification of the preferred vector field as $\dot{x}^\al$ in \S\ref{dos}
remains the same as the stresses $T^{\mu\nu}$ \ref{2.1.27}, \ref{2.1.46},  \ref{n39}
and hence $T^{\mu\nu}_{.~.~;\nu}$ are unaltered.    
Thus under the transformation $\ta\rightarrow\ta'$ the preferred vector field 
just transforms as 
$\fr{\rm{d}x^\al}{\rm{d}\ta}\rightarrow
\fr{\rm{d}\ta'}{\rm{d}\ta}\fr{\rm{d}x^\al}{\rm{d}\ta'}$.

Having identified the vector $\dot{x^\al}$, rather than say $P^\al$,  
as the preferred one,  
one can use the standard decomposition \cite{HE} and the appendix \S\ref{app} 
to calculate the rotation and shear.
\section{The String.}\lb{string}
A point particle can be thought of as a minimal surface 
of zero dimension $p=0$,  
and a string a minimal surface of dimension one $p=1$,
and a membrane a minimal surface of arbitrary $p<d$ the dimension of the spacetime.
In this section the simplest string action and in the next section
the simplest membrane action are investigated.
\subsection{The Nambu-Goto Action.}\lb{nga}
The action of the string [Eq.I.16 of \cite{scherk}] 
can be taken in a form that corresponds to \ref{2.1.1}
\be
S=\int^{\ta_2}_{\ta_1}\int^\pi_0\dt\ds\Lg,
\lb{3.1.1}
\ee
where the standard specific Lagrangian is the Nambu-Goto Lagrangian
\be
\Lg=-\fr{\ar}{2\pi\al'},
\lb{3.1.2}
\ee
the area $\ar$ \ref{3.1.3} 
is a generalization of the lenght $\el$ \ref{2}
\be
\ar\equiv\sq{(\dot{x}\cdot x')^2-\dot{x}^2x'^2},   
\lb{3.1.3}
\ee
also in \ref{3.1.1}, $\ta$ is the evolution parameter,
$\si$ is the kinematic parameter,
$\al'$ is the string tension,
$\dot{x}^\al=\rm{D}x^\al/\rm{d}\ta$, 
$x'=\rm{D}x^\al/\rm{d}\si$.
The velocities $\dot{x}$ and $x'$ can be varied in a similar manner
to \ref{2.1.2} to give the varied action
\ber 
\de S=-\int^{\ta_2}_{\ta_1}\rm{d}\ta\int^\pi_0\ds\de x^\mu
\left[\Dt\fr{\p\Lg}{\p\dot{x}^\mu}
     +\Ds\fr{\p\Lg}{\p x'^\mu}   
     \right]\nonumber\\
+\int^\pi_0\ds
 \fr{\p\Lg}{\p \dot{x}^\mu}\de x^\mu|^{\ta_2}_{\ta_1} 
+\int^{\ta_2}_{\ta_1}\dt
 \fr{\p\Lg}{\p x'^\mu}\de x^\mu|^{\si=\pi}_{\si=0}.
\lb{3.1.4}
\ear
Choosing initial and final positions on the string to be fixed so
that \ref{2.1.5} is obeyed the second term vanishes.  
For open strings the third term vanishes when the edge condition
{\it c.f.} eq.I.18 of \cite{scherk}
\be
\fr{\p\Lg}{\p x'^\mu}|_{\si=0}
=\fr{\p\Lg}{\p x'^\mu}|_{\si=\pi}
\lb{3.1.5}
\ee
is obeyed.
For closed strings $x_\mu(\ta,\si+2\pi)=x_\mu(\ta,\si)$
{\it c.f.} \S II.7 of \cite{scherk}.
The vanishing of the first term gives the string equation of motion
\be
 \Dt\fr{\p\Lg}{\p \dot{x}^\mu}
+\Ds\fr{\p\Lg}{\p x'^\mu}=0
\lb{3.1.6}
\ee
\subsection{Reduction to the Point Particle.}\lb{rpp}
The theory of the string,  a $p=1$ minimal surface,
must somehow incorporate that of  a $p=0$ minimal surface,
which describes a point particle.
To see how use the reduction equation
\be
\fr{\p\Lg}{\p x'^\al}=0,
\lb{3.1.7}
\ee
the string equation of motion 
\ref{3.1.6} 
takes the same form as 
\ref{2.1.6}.
For the specific Lagrangian \ref{3.1.2}, the reduction equations
\be
\dot{x}\cdot x'\st{\ref{3.1.2}}{=}0
~~~\&~~~
x'^2\st{\ref{3.1.2}}{=}+1,
\lb{3.1.8}
\ee
reduce the area $\ar$ 
\ref{3.1.3} 
to the lenght $\el$ 
\ref{2}.
The reduction equation
\be
m=\fr{1}{2\pi\al'}
\lb{3.1.9}
\ee
equates the coupling constants so that the point particles 
equation of motion \ref{2.1.7} is recovered.
\subsection{Introducing Momenta.}\lb{inmm}
One can define momenta
\be
P^\ta_\mu
\equiv\fr{\de S}{\de \dot{x}^\mu}
=\fr{\p {\cal L}}{\p\dot{x}^\mu},~~~    
P^\si_\mu
\equiv\fr{\de S}{\de x'^\mu}
=\fr{\p {\cal L}}{\p x'^\mu},
\lb{3.1.10}
\ee
note that $\ta\&\si$ are superscripted not subscripted.
These allow the edge condition \ref{3.1.5} to be put in the form
\be
P^{\mu\si}|_{\si=0}=P^{\mu\si}|_{\si=\pi},
\lb{3.1.11}
\ee
and the equations of motion \ref{3.1.6} to be put in the simple form
\be
\dot{P}^{\ta\mu}+P'^{\si\mu}=0,
\lb{3.1.12}
\ee
where $\dot{P}^{\ta\mu}=\rm{D}P^{\ta\mu}/\rm{d}\ta$
and   $P'^{\si\mu}=\rm{D}P^{\si\mu}/\rm{d}\si$.
For the specific Lagrangian \ref{3.1.1} the momenta \ref{3.1.10} are
\be
-2\pi\al'\ar P^{\ta\mu}
\st{\ref{3.1.2}}{=}\dot{x}\cdot x'x'^\mu-x'^2\dot{x}^\mu,~~~~~~
-2\pi\al'\ar P^{\si\mu}
\st{\ref{3.1.2}}{=}\dot{x}\cdot x'\dot{x}^\mu-\dot{x}^2x'^\mu,
\lb{3.1.13}
\ee
and using \ref{3.1.8} the reduction equation \ref{3.1.7} becomes
\be
\dot{x}^2x'^\mu=0,
\lb{3.1.14}
\ee
and $P^{\ta\mu}$ reduces to $P^\mu$ \ref{2.1.8} 
and $P^{\si\mu}$ vanishes.
Inverting \ref{3.1.13}
\ber
\ar\dot{x}^\mu
=-2\pi\al'(\dot{x}^2P^{\ta\mu}+\dot{x}\cdot x'P^{\si\mu})
=2\pi\al'(P^{\si^2} P^{\ta\mu}-P^\ta\cdot P^\si P^{\si\mu}),\n
\ar x'^\mu
=-2\pi\al'(x'^2P^{\si\mu}+\dot{x}\cdot x'P^{\ta\mu})
=2\pi\al'(P^{\ta^2} P^{\si\mu}-P^\ta\cdot P^\si P^{\ta\mu}).
\lb{3.1.15}
\ear
\subsection{Introducing the Hessian.}\lb{hes}
The specific projection tensor is defined by
\ber 
h^{\mu\nu}&\st{\ref{3.1.2}}{\equiv}&g^{\mu\nu}+\fr{1}{\ar^2}\times             
\left(\dot{x}^2x'^\mu x'^\nu +x'^2\dot{x}^\mu\dot{x}^\nu
-(\dot{x}\cdot x')(\dot{x}^\mu x'^\nu+x'^\mu\dot{x}^\nu)\right),\n
h&=&d-2.
\lb{3.1.16}
\ear
this projection tensor is a generalization 
of the familiar one in general relativity \ref{2.1.12}.
It can be written in several forms using momenta only or, 
momenta and coordinates, the most convenient is
\be
h^{\mu\nu}\st{\ref{3.1.2}}{=}g^{\mu\nu}+\fr{2\pi\al'}{\ar}
\times(P^{\ta\mu}\dot{x}^\nu+P^{\si\mu} x'^\nu),
\lb{3.1.17}
\ee
which corresponds to the middle equality of the first line of \ref{2.1.12}.
The projection tensor \ref{3.1.16} obeys the equations
\ber
h^{\mu\nu}\dot{x}^{*}_\nu&=&\dot{x}^{*\mu}+\fr{1}{\ar^2}
\times\left((\dot{x}^2x'\cdot\dot{x}^{*}-\dot{x}\cdot x'\dot{x}\cdot\dot{x}^{*})x'^\mu
+(x'^2\dot{x}\cdot\dot{x}^{*} -\dot{x}\cdot x' x'\cdot\dot{x}^{*})\dot{x}^\mu\right),\n
h^{\mu\nu}\st{*}{x}_\nu&=&0,~~~                        
h=d-2,~~~
h^{\mu\al}h_{\nu\al}=h^\mu_{.~\nu},
\lb{3.1.18}
\ear
with $*=0~ \rm{or}~ '$ and $h^{\mu\nu}x"$ following by symmetry.
The specific projection tensor allows the derivatives 
of the momenta to be expressed as
\be
2\pi\al'\ar P^{\ta*\mu}
=h^{\mu\nu}(x'^2\dot{x}^*_\nu-\dot{x}\cdot x'x'^{*}_\nu)
 \pm\dot{x}\cdot x'^{*}x'^\mu
 \mp x'\cdot x'^{*}\dot{x}^\mu
\lb{3.1.19}
\ee
with the top sign in the last two terms for $*=0$
and the bottom sign for $*='$,
and similarly for $P^{\si\mu}$.
The Hessian is
\ber
W^{\ta\ta\mu\nu}&\st{\ref{3.1.2}}{=}&\fr{x'^2h^{\mu\nu}}{2\pi\al'\ar},~~~
W^{\si\ta\nu\mu}\st{\ref{3.1.2}}{=}-\fr{1}{2\pi\al'\ar}
   (\dot{x}\cdot x'h^{\mu\nu}+\dot{x}^\nu x'^\mu-x'^\nu\dot{x}^\mu)\n
W^{\si\si\mu\nu}&\st{\ref{3.1.2}}{=}&\fr{\dot{x}^2h^{\mu\nu}}{2\pi\al'\ar},~~~
W^{\ta\si\nu\mu}\st{\ref{3.1.2}}{=}-\fr{1}{2\pi\al'\ar}
   (\dot{x}\cdot x'h^{\mu\nu}+\dot{x}^\mu x'^\nu-\dot{x}^\nu x'^\mu)
\lb{3.1.20}
\ear
the equations of motion \ref{3.1.12} become
\ber
2\pi\al'\ar(\dot{P}^{\ta\mu}+P'^{\si\mu})&\st{\ref{3.1.2}}{=}&
2\pi\al'\ar\left(\ddot{x}_\nu W^{\ta\ta\mu\nu}
     +x''_\nu W^{\si\si\mu\nu}
     +\dot{x}'_\nu(W^{\ta\si\mu\nu}+W^{\si\ta\mu\nu})\right)\n 
&\st{\ref{3.1.2}}{=}&h^{\mu\nu}(x'^2\ddot{x}_\nu
+\dot{x}^2x''_\nu-2\dot{x}\cdot x' \dot{x}'^\nu)=0,
\lb{3.1.21}
\ear
\subsection{The Stress and Its Derivatives.}\lb{sid}
The energy-momentum \ref{2.1.22} for \ref{3.1.2} is
\be
T_{\mu\nu}=-\fr{\ar}{2\pi\al'}(2h_{\mu\nu}-g_{\mu\nu}),~~~~~~~~~~
T=\fr{(4-d)\ar}{2\pi\al'},
\lb{3.1.23}
\ee
and \ref{3.1.23} is traceless in four dimensions.
At first sight, using \ref{2.1.27b} for the perfect fluid equivalence of this 
energy-momentum one has
\be
p=\mu=\fr{\ar}{2\pi\al'}
\lb{3.1.23b}
\ee
however this assumes that the $h^{\mu\nu}$'s  \ref{2.1.12} and \ref{3.1.16} 
are the same which they are not.
Choosing two interacting fluids with stress
\be
T^{\mu\nu}=(\mu_1+p_1)V^\mu V^\nu 
+(\mu_2+p_2)W_\mu W_\nu 
+(p_1+p_2)g^{\mu\nu}
+2\xi V_{(\mu}W_{\nu)}
\lb{3.1.23c}
\ee
and choosing non-unit vectors $V^\mu=\dot{x}^\mu/\ar$,  $W^\mu=x'^\mu/\ar$ 
the interaction parameter is
\be
\xi=(\dot{x}\cdot x')\fr{\ar}{\pi\al'}
\lb{3.1.23d}
\ee
and there is one free function among the four $\mu_1,\mu_2,p_1,p_2$
letting it be $p_2$ gives
\be
p_1=-\fr{\ar}{2\pi\al'}-p_2,~~~
\mu_1=p_2+(1-2x'^2)\fr{\ar}{2\pi\al'},~~~
\mu_2=-p_2-\fr{\dot{x}^2\ar}{\pi\al'}.
\lb{3.1.23e}
\ee
For investigation of the stress's derivatives it is easiest to work
with the mixed form of the projection tensor \ref{3.1.17} giving
\be
T^{\mu\nu}=-2(P^{\ta\mu}\dot{x}^\nu+P^{\si\mu} x'^\nu)-\fr{\ar g^{\mu\nu}}{2\pi\al'}.
\lb{3.1.24}  
\ee
Stress conservation \ref{2.1.23} gives
\be
T^{\mu\nu}_{.~.~;\nu}=-\fr{\ar^\mu}{2\pi\al'}
-2(\dot{P}^{\ta\mu}+P'^{\si\mu}+\st{\ta}{\Te}P^{\ta\mu}+\st{\si}{\Te}P^{\si\mu})
\lb{3.1.25}  
\ee
where the absolute derivatives are taken to be given by
\be
\Dt=\dot{x}^\nu\na_\nu,~~~
\Ds=x'^\nu\na_\nu,
\lb{3.1.26}
\ee
and the expansions are
\be
\st{\ta}{\Te}=\dot{x}^\nu_{.~\nu},~~~
\st{\si}{\Te}=x'^\nu_{.~\nu}.
\lb{3.1.27}
\ee
After using the equations of motion \ref{3.1.12} stress conservation gives
\be
\ar^\mu=-4\pi\al'(\st{\ta}{\Te}P^{\ta\mu}+\st{\si}{\Te}P^{\si\mu})
\lb{3.1.28}
\ee
which equates the expansion of $x$ to the change in the area,
compare \ref{2.1.28}.
Equations \ref{3.1.27} suggest that there are two preferred vector fields 
$\dot{x}^\mu$ and $x'^\mu$,
however $\ta$ and $\si$ are not unique and can transform,  so that mixtures of
$\dot{x}^\mu$ and $x'^\mu$ might be better preferred vector fields.
The best way to approach this is through vector notation with $a,b\dots=\ta,\si$,
this can be achieved with not much more effort by using the Dirac membrane.
\section{The Membrane.}\lb{mem}
\subsection{The Action.}\lb{act}
The membrane action is 
\be
S_D=\int_M\rm{d}^{p+1}\xi\sq{-\ga}\Ls,~~~
\ga_{ab}=g_{\mu\nu}\p_a x^\mu\p_b x^\nu,~~~
\sq{-\ga}=(-\det \ga_{ab})^\fr{1}{2},
\lb{4.1}
\ee
where $\Lg=\sq{-\ga}\Ls$.
Choosing
\ber
p&=&1,~~~
\Ls=k=-\fr{1}{2\pi\al'},~~~
a,b\dots=\ta,\si,~~~
\rm{d}^2\xi=\dt\ds,\\
\ga&=&\det(\ga_{ab})=-\ar^2,~~~
\ga_{ab}=\left(\begin{array}{cc}\dot{x}^2&\dot{x}\cdot x'\\
                                x'\cdot\dot{x}&x'^2\\
               \end{array}\right),~~~
\ga\ga^{ab}=\left(\begin{array}{cc}x'^2&-x'\cdot\dot{x}\\
                                   -\dot{x}\cdot x'&\dot{x}^2
               \end{array}\right),
\nonumber
\lb{4.2}
\ear
this reduces to the Nambu-Goto action \ref{3.1.2}.
The first fundamental form is defined by
\be
\ff^{\mu\nu}\equiv\ga^{ab}x^\mu_ax^\nu_b,~~~
\ff^{\mu\rh}\ff^\nu_{.\rh}=\ff^{\mu\nu},~~~
\ff^\rh_{.\rh}=\ga^c_{.c}=x^{\rh c}_{..}x_{\rh c}=p+1,
\lb{4.2b}
\ee
which allows the generalization of the projection tensors 
\ref{2.1.12} and \ref{3.1.16} to be expressed as
\be 
h^{\mu\nu}\st{\ref{4.1}}{=}g^{\mu\nu}-\ff^{\mu\nu},~~~
h=d-1-p.
\lb{4.3}
\ee
Choosing
\be
\Ls_D=k,~~~
\Ls_P=\fr{k}{2}\left\{\ff^\rh_{.\rh}-(p-1)\right\},~~~
\Ls_C=k\left\{\fr{\ff^\rh_{.\rh}}{p+1}\right\}^\fr{p+1}{2},~~~
\Ls=k\Lf(\ff),
\lb{4.3b}
\ee
gives the Dirac \cite{dirac} or determinant form which generalizes \ref{2}, 
the second order form generalizing \ref{2.1.49}, and conformal actions respectively.
Rather than work with these three Lagrangians seperately 
it is often easier to work with $\Lf$, 
where $\Lf(\ff)$ is a function of $\ff^\rh_\rh$,
as it includes all of them as special cases.
Substituting $\ff^\rh_\rh=p+1$ into the first three of these actions gives
$\Ls_D=\Ls_P=\Ls_C=k$

The spacetime metric stress \ref{2.1.22} 
generalizing \ref{2.1.27} and \ref{3.1.21} is
\ber 
T^{\mu\nu}_\ff&=&k\sq{-\ga}\left\{((p+1)\Lf+2\Lf')h^{\mu\nu}
                          -(p\Lf+2\Lf')g^{\mu\nu}\right\},\n
T&=&k\sq{-\ga}\left\{(d-(p+1)^2)\Lf-2(p+1)\Lf'\right\},
\lb{4.4}
\ear
where $\Lf'=\p\Lf/\p\ff^\rh_\rh$.
This can be traceless for example for $\Lf=(\ff^\rh_\rh)^n$ with $2n=d-(p+1)^2$
The stresses for the first three lagrangians of \ref{4.3b} are
\be
T^{\mu\nu}_D=k\sq{-\ga}\left\{(p+1)h^{\mu\nu}-pg^{\mu\nu}\right\},~~~
T^{\mu\nu}_P=T^{\mu\nu}_C=k\sq{-\ga}\left\{(p+2)h^{\mu\nu}-(p+1)g^{\mu\nu}\right\}.
\lb{4.4b}
\ee
The reason that $T^{\mu\nu}_P$ and $T^{\mu\nu}_C$ are different from $T^{\mu\nu}_D$
is that the contribution to the stress from the constraints has not been taken into 
account,  see equation \ref{2.1.46} for how this is done for the point particle.
The stresses calculated with the internal metric are described in the next section \S\ref{wsr}.
The spacetime metric stress \ref{4.4} can be projected
\be
h_{\mu\rh}T^{\rh\nu}=k\sq{-\ga}\Lf h^\nu_{.\mu},~
x_{\nu a}T^{\mu\nu}=-k\sq{-\ga}(p\Lf+2\Lf')x^\mu_{.a}.
\lb{4.7}
\ee
The spacetime metric stress \ref{4.4} is of the form of functions times $g_{\mu\nu}$
and $h_{\mu\nu}$ and so has a fluid anology,
assuming that there is a fluid of a form generalizing \ref{2.1.27b}
\be
T^{\mu\nu}=(\mu+\pr)h^{\mu\nu}-\mu g_{\mu\nu},
\lb{4.7b}
\ee
which now has a much more complex projection tensor $h_{\mu\nu}$ given by \ref{4.3}.
\ref{4.4} gives the pressure and density
\be
\pr=k\sq{-\ga}\Lf,~~~
\mu=k\sq{-\ga}(p\Lf+2\Lf'),
\lb{4.7c}
\ee
and this has a $\Ga$-equation of state
\be
\pr=(\Ga-1)\mu,
\lb{4.7d}
\ee
with
\be
\Ga=\fr{(p+1)\Lf+2\Lf'}{p\Lf+2\Lf'},~~~p\Lf+2\Lf'\ne0,
\lb{4.7e}
\ee
the $p=0$ case is more easily treated seperately,  see \ref{1.1.27d}.
For $p\ne0$ the projection tensor involves more than one vector field and so does
not have a familiar general relativistic interpretation of being orthogonal to a
moving observer.  
The $\Ga$-equation of state usually has $\Ga$ a number rather than a function,
in the present case this can be achieved using $\Lf=(\ff^\rh_\rh)^n$ giving
\be
\Ga=\fr{(p+1)^2+2n}{p(p+1)+2n},~~~
n=\fr{(p+1)(p\Ga-(p+1))}{2(1-\Ga)}.
\lb{4.7f}
\ee
In general a system described by a set of vectors has a fluid anology when there exists
an internal metric so that a surface with first fundamental form and projection can be defined.

The equation of motion resulting from varying \ref{4.1} with respect to $x$ is
\be
-k\p_a\left\{\sq{-\ga}x^{\mu a}(\Lf+2\Lf')\right\}=0.
\lb{4.9}
\ee
Substituting $\Lg_P$ gives a constant times Polchinski \cite{polchinski}eq.1.2.25.
From now on assume $\Lf+2\Lf'\ne0$.
The momenta are
\be
P^{\mu a}\equiv\fr{\de S}{\de\p_a x_\mu}
=\fr{\p \Lg}{\p\p_ax_\mu}
=k\sq{-\ga}(\Lf+2\Lf')x^{\mu a}
\lb{4.10}
\ee
so that the equation of motion generalizing \ref{3.1.12} is
\be
\na_aP^{\mu a}=0.
\lb{4.11}
\ee
In terms of the momenta \ref{4.10} 
the first fundamental form \ref{4.2b} is
\be
\ff^{\mu\nu}=\fr{P^\mu_ax^{\nu a}}{\sq{-\ga}(\Lf+2\Lf')}
\lb{4.11b}
\ee
from this the projection tensor \ref{4.3} can be constructed and it
generalizes \ref{3.1.17}.
The stress \ref{4.4} can now be expressed as
\be
T^{\mu\nu}=\Lg g^{\mu\nu}-FP^{\mu a}x^\nu_a,~~~
F\equiv\fr{p\Lf+2\Lf'}{\Lf+2\Lf'}\Ga
\lb{4.14}
\ee
with $\Ga$ given by \ref{4.7e}
and this generalizes \ref{3.1.24}.
\subsection{Derivatives of the Stress.}
Covariantly differentiating with respect to $x^\nu$ gives the generalization of \ref{3.1.25}
\be
T^{\mu\nu}_{.~.~;\nu}=\Lg^\mu-(FP^{\mu a}x^\nu_a)_\nu,
\lb{4.15}
\ee
where the absolute derivatives are
\be
\na_a=\p_ax^\nu\na_\nu,
\lb{4.16}
\ee
and the $p+1$ expansions are
\be
\st{a}{\Te}=\p_a x^\nu_{.\nu}.
\lb{4.17}
\ee
After using the equations of motion \ref{4.11} stress conservation gives
\be
\Lg^\mu=P^{\mu a}\left(F\st{a}{\Te}+F_a\right),
\lb{4.18}
\ee
for the Dirac Lagrangian $F=p+1$ and the second term on the right hand side vanishes.
This equation \ref{4.18} equates the spacetime derivative of the Lagrangian to 
the expansions of $x^\mu$,
this generalizes \ref{3.1.28} and \ref{2.1.37e}.
From \ref{4.17} and \ref{4.18} it is apparent that the best vector field 
for calculating shear and vorticity is
\be
V^\mu=\sum^p_{a=1}\p_ax^\mu.
\lb{4.19}
\ee
using this vector one can use the standard decomposition \cite{HE}, 
see also the appendix \S\ref{app} below,
to calculate the rotation and shear of a membrane.
\section{Worldsheet Restriction.}\lb{wsr}
Define the internal metric stress
\be
T^{ab}\equiv\fr{1}{\sq{-\ga}}\fr{\de S}{\de \ga_{ab}}
=-2\fr{\p \Ls}{\p\ga_{ab}}+\ga^{ab}\Ls
\lb{5.2b}
\ee
then the restricted spactime stress is
\be
\bar{T}^{\mu\nu}=x^\mu_ax^\nu_bT^{ab}
\lb{5.2c}
\ee
compare the equations Carter \cite{carter}eq.6.3.
$\bar{T}^{\mu\nu}$ is part of $T^{\mu\nu}$,
unfortunately there does not seem to be a way of
finding it using projections.
Let $n_\nu$ be any normal to the membrane such that
\be
n_\mu x^\mu_a=0,
\lb{5.2d}
\ee
then from \ref{5.2c}
\be
n_\mu \bar{T}^{\mu\nu}=0,
\lb{5.2e}
\ee
implying that the stress is restricted to the surface, 
and so is only part of $T^{\mu\nu}$.
Any projection tensor constructed from $n_\mu$ is not the same as \ref{4.3}.

The restricted spacetime stress can be thought of in the following way.
Taking the lagrangian corresponding to \ref{4.1} to be $\bar{\Lg}$
the restricted action is
\be
S=\int d^{p+1}\xi\sq{-\ga}\bar{\Lg}(\xi)
\lb{5.1}
\ee
with $\ga$ being given by the second equation of \ref{4.1}.
In this case $T^{\mu\nu}$ contains a factor of a $(d-p)$-dimensional
$\de$-function ensuring that it vanishes for all values of $x$ 
not on the worldsheet.
What multiplies the $\de$-function ia a quantity $\bar{T}^{\mu\nu}$,
which can be defined by requiring that the change of the action
under a small change of the metric is
\be
\de S=-\fr{1}{2}\int d^{p+1}\xi\sq{-\ga}\bar{T}^{\mu\nu}\de g_{\mu\nu}(X(\xi)),
\lb{5.2}
\ee
compare Carter \cite{carter}eq.6.2.
and this variation again gives \ref{5.2c}.

For the lagrangians \ref{4.3b} 
\ber
T^{ab}_D&=&k\ga^{ab},~~~
T^{ab}_\ff=-2k\Lf'x^a_\rh x^{\rh b}+k\Lf\ga^{ab},~~~
\bar{T}^{\mu\nu}_\ff=k(\Lf-2\Lf')\ff^{\mu\nu}\n
T^{ab}_P&=&-\fr{k}{2}\left(2x^a_\rh x^b_\rh-\ga^{ab}(\ff^\rh_\rh-p+1)\right)\n
T^{ab}_C&=&\fr{k}{p+1}\left(\fr{\ff^\rh_\rh}{p+1}\right)^\fr{p-1}{2}
              \left(-(p+1)x^{\rh a}x^b_\rh+\ff^\rh_\rh\ga^{ab}\right)
\lb{5.3}
\ear
The second of these is Polchinski \cite{polchinski}eq.1.2.22 when $p=1,k=1/\al'$.
The traces of \ref{5.3} are
\be
T^a_{D a}=k(p+1),~~~
T^a_{P a}=T^a_{C a}=0.
\lb{5.4}
\ee
After substituting $\ff^\rh_\rh=p+1$ the conservation equations are
\be
T^{ab}_{D~|b}=0,~~~
T^{ab}_{P~|b}=T^{ab}_C=-k(x^a_\rh x^{\rh b})_a,~~~
\lb{5.5}
\ee
In this case as the derivatives are with respect to the internal metric
so there is no resulting spacetime vector field to use to construct spacetime
rotation and shear. 
\section{Appendix}\lb{app}
A vector field can be decomposed
\be
V_{\mu;\nu}
=\om_{\mu\nu}+\si_{\mu\nu}+\fr{1}{3}\Te h_{\mu\nu} \dot{V}_\mu V_\nu,
\lb{A1}
\ee
where the vorticity tensor and vector are
\be
\om_{\mu\nu}=h^\al_\mu h^\bt_\nu V_{[\al;\bt]},~~~
\om^\mu=\fr{1}{2}\et^{\mu\nu\al\bt}V_\nu\om_{\al\bt},
\lb{A3}
\ee
the expansion tensor and scalar are
\be
\Te_{\mu\nu}=h^\al_\mu h^\bt_\nu V_{(\al;\bt)},~~~
\Te=V^\mu_{.~;\mu},
\lb{A5}
\ee
and the shear is
\be
\si_{\mu\nu}=\Te_{\mu\nu}-\fr{1}{3}h_{\mu\nu}\Te,
\lb{A6}
\ee
\newpage
\section{Conclusion.}
The vector \ref{4.19}, rather than say  $\sum^p_{a=1}P^{\mu a}$,
produces conservation equations \ref{4.15},
terms of which can be removed by the equations of motion,
leaving a simple relationship between the spacetime derivative 
of the Lagrangian and the expansion of \ref{4.19}:
thus it is best for forming geometric objects,
such as rotation and shear.

\section{Acknowledgement}
I would like to thank Tom Kibble for reading the paper 
and pointing out the references of Carter.

\end{document}